# Angular Upsampling in Infant Diffusion MRI Using Neighborhood Matching in x-q Space

**Geng Chen** [1], **Bin Dong** [2], **Yong Zhang** [3], **Weili Lin** [1], **Dinggang Shen** [1,4,*] **and Pew-Thian Yap** [1,*]

[1] *Department of Radiology and Biomedical Research Imaging Center (BRIC), University of North Carolina at Chapel Hill, NC, U.S.A.*
[2] *Beijing International Center for Mathematical Research, Peking University, Beijing, China*
[3] *Colin Artificial Intelligence Lab, Richmond, Canada*
[4] *Department of Brain and Cognitive Engineering, Korea University, Seoul, Korea*

Correspondence*:
Pew-Thian Yap and Dinggang Shen
ptyap@med.unc.edu and dgshen@med.unc.edu

## ABSTRACT

Diffusion MRI requires sufficient coverage of the diffusion wavevector space, also known as the $q$-space, to adequately capture the pattern of water diffusion in various directions and scales. As a result, the acquisition time can be prohibitive for individuals who are unable to stay still in the scanner for an extensive period of time, such as infants. To address this problem, in this paper we harness non-local self-similar information in the $x$-$q$ space of diffusion MRI data for $q$-space upsampling. Specifically, we first perform neighborhood matching to establish the relationships of signals in $x$-$q$ space. The signal relationships are then used to regularize an ill-posed inverse problem related to the estimation of high angular resolution diffusion MRI data from its low-resolution counterpart. Our framework allows information from curved white matter structures to be used for effective regularization of the otherwise ill-posed problem. Extensive evaluations using synthetic and infant diffusion MRI data demonstrate the effectiveness of our method. Compared with the widely adopted interpolation methods using spherical radial basis functions and spherical harmonics, our method is able to produce high-resolution diffusion MRI data with greater quality, both qualitatively and quantitatively.

## 1 INTRODUCTION

Infant brain development involves complex cerebral growth and maturation with the white matter (WM) undergoing rapid myelination and synaptogenesis (Qiu et al., 2015). Diffusion MRI (DMRI) has been widely employed to study this developmental process *in vivo* (Yap et al., 2011; Huang et al., 2013; Dubois et al., 2014; Qiu et al., 2015). For instance, using diffusion tensor imaging, researchers have observed an increase in the fractional anisotropy (FA) during the first few years of life (Dubois et al., 2014; Qiu et al., 2015), implying more restriction on water movement owing to the ensheathment of oligodendrocytes around the axons. Mean diffusivity (Dubois et al., 2014; Qiu et al., 2015) and structural connectivity (Yap et al., 2011; Huang et al., 2013) have also been used to study early brain development.

In practice, a DMRI protocol involves acquiring multiple diffusion-weighted (DW) images, each corresponding to a wavevector **q** in $q$-space. The vector **q** can be separated into a scalar wavenumber $|\mathbf{q}|$ and a diffusion encoding direction $\hat{\mathbf{q}} = \mathbf{q}/|\mathbf{q}|$. The effects of both diffusion time, $t$, and wavenumber are summarized using a quantity called $b$-value, which is defined as $b = t|\mathbf{q}|^2$. In general, the number of DW images increases with the complexity of the diffusion model one is interested in fitting to the data. For example, the diffusion tensor model requires a minimum of only 6 DW images and 1 non-DW image. However, for a more realistic account of the white matter (WM) neuronal architecture, such as fiber crossings and intra-/extra-cellular compartments, more DW images are typically needed to cater to more sophisticated models (Yap et al., 2016b; Ning et al., 2015; Ye et al., 2016).

In practice, the window of opportunity for imaging infants is short. To put this in perspective, in the Human Connectome Project (HCP) (Van Essen et al., 2012) each individual was allotted a DMRI scan time of about an hour. However, in the Baby Connectome Project (BCP) (Fallik, 2016; Cao et al., 2017), the tolerable scan time is well below 15 minutes. Infants are typically scanned without sedation while they are asleep. The scanning may also need to be terminated prematurely if the infant is awakened by the loud acoustic noise and sudden vibrations caused by the rapid switching of gradient amplitude and polarity (Hutter et al., 2017; McJury and Shellock, 2000). The short acquisition time precludes a denser coverage of $q$-space, limiting studies to simpler models such as the diffusion tensor.

One way to increasing the angular resolution without additional acquisition time is by post-acquisition upsampling. Currently, few methods have been developed for this purpose. The most commonly used method is interpolation using spherical radial basis functions (SRBFs) (Tuch, 2004). An alternative solution is interpolation using spherical harmonics (SHs) (Descoteaux et al., 2007), which first performs spherical decomposition and then recovers the expected DMRI data using SHs and the corresponding coefficients. However, both SRBF interpolation and SH interpolation only employ $q$-space measurements for upsampling and neglect signal correlation across voxels in $x$-space. In this paper, we introduce a method for $q$-space upsampling by leveraging non-local self-similar information in the $x$-$q$ space. Our consideration of the joint $x$-$q$ space allows information from curved WM structures to be used more effectively for upsampling. The contributions of the current work are as follows: (1) Upsampling based directly on the DW images. No model is assumed, other than non-local smoothness; (2) Non-local relationships of $q$-space samples are learned over both $x$-space and $q$-space to help regularize the otherwise ill-posed inverse problem involved in recovering the high $q$-space resolution data from the low-resolution counterparts; (3) Our method can effectively use information from curved WM structures because joint $x$-$q$ space neighborhood matching is tolerant of angular differences; (4) Our results indicate that the proposed method can help compensate for insufficient sampling in $q$-space due to time constraints.

Part of this work has been reported in our recent conference paper (Chen et al., 2017b). Herein, we present (1) New insights into the application of our method to infant DMRI data; (2) More detailed descriptions of the proposed method; (3) More complete mathematical derivation details; (4) Comparison with the state-of-the-art method, SH interpolation; (5) New experimental results on synthetic data and infant data acquired at different time points. None of these is part of the conference publication.

The rest of the paper is organized as follows. In Section 2, we give a detailed description of the proposed method. In Section 3, we demonstrate the effectiveness of the method with both synthetic and real infant DMRI data. In Section 4, we provide further discussion on this work. Finally, in Section 5, we conclude this work.

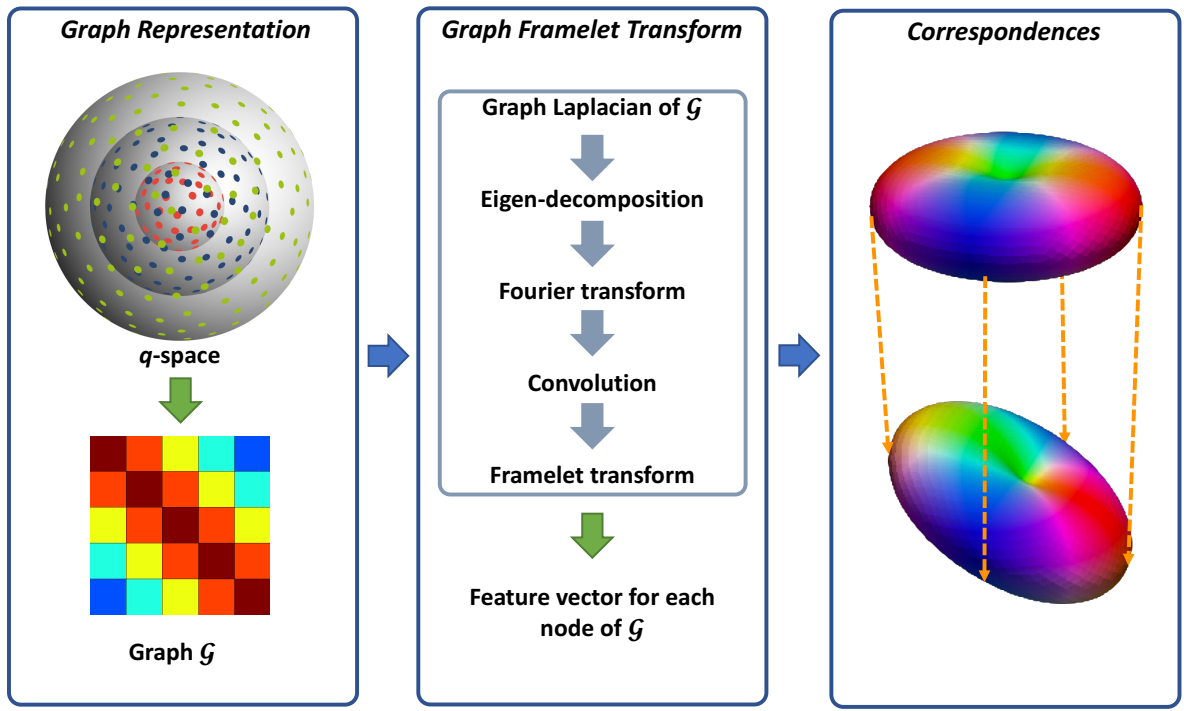

**Figure 1. Overview.** (Left) Representing the $q$-space sampling domain using a graph with affinity matrix determined by kernels for diffusion gradient directions and strengths. (Middle) Feature computation using GFTs. (Right) Neighborhood matching.

## 2 METHODS

Each signal in the $x$-$q$ space is associated with a voxel location in the $x$-space and a set of diffusion gradient direction and strength in the $q$-space. The signals are typically collected via a set of DW images, each corresponding to a point in $q$-space. Our approach first establishes the relationships of the signal samples in $x$-$q$ space using robust neighborhood matching. These relationships are then used to regularize the inverse problem that needs to be solved to recover the high-resolution data.

### 2.1 Signal Correspondences in $x$-$q$ Space

A similarity weight is assigned to each pair of signal measurements in $x$-$q$ space based on a robust patch matching method similar to (Chen et al., 2016, 2017a). Conventional $x$-space neighborhood matching methods, such as non-local means, has demonstrated remarkable performance in locating recurrent image patterns. Our work uses an extension of this to $x$-$q$ space so that neighborhood matching can be performed effectively for curved WM structures. In what follows, we discuss how we generate a feature vector for each point in $x$-$q$ space and how we perform neighborhood matching for each pair of points in $x$-$q$ space. See Fig. 1 for an overview.

2.1.1 Graph representation for $q$-space

As illustrated in Fig. 1, we use a graph $\mathcal{G}$ to represent the $q$-space. For this purpose, we compute the adjacency weight between each two nodes in $q$-space using two kernels for diffusion gradient directions and strengths. In this way, the geometric relationships in $q$-space are encoded in the graph $\mathcal{G}$ where a large weight indicates two nodes sharing similar gradient directions and diffusion weightings.

### 2.1.2 Feature Computation

We use graph framelet transforms (GFTs) (Dong, 2017; Yap et al., 2016a) to generate a multi-scale feature vector for each point in $x$-$q$ space. This is achieved by viewing each point in $q$-space at each location in $x$-space as a node of a graph and generating features for each node using GFTs for correspondence matching. GFTs can be used to generate a wavelet representation for data that are not necessarily residing on a Cartesian grid. In our case, the data in $q$-space is in general unstructured in the sense that the sampling points may not distribute with perfect uniformity. GFTs involve computing the eigenspectrum based on the graph Laplacian, and then performing multiresolution analysis based on the spectrum. Note that GFTs can in fact be viewed as convolutions on the graph and are hence features of local neighborhoods. GFT features can therefore be used for neighborhood matching. The interested reader is referred to (Dong, 2017) for more details.

### 2.1.3 Neighborhood Matching in $x$-$q$ Space

At each location in $x$-space, the $q$-space signal samples can be seen as a function defined on the nodes of a graph. Using the computed GFT features, dense node-to-node correspondence can be performed with precision and efficiency, within a graph (i.e., within a voxel) or across graphs (i.e., across voxels). As a result of the correspondence matching, a similarity weight is assigned to each pair of nodes. For more details, please refer to (Chen et al., 2016, 2017a). Our method is akin to the spectral matching method described in Lombaert et al. (2013), but allows a richer characterization of each node using frames.

## 2.2 Angular Upsampling

We use the $x$-$q$ space data relationships determined in the previous section to guide data upsampling in the $q$-space. In general, recovering the high angular resolution (HAR) data from the low angular resolution (LAR) data is an ill-posed inverse problem. Solution can however be feasible by imposing structure via harnessing prior information to reduce the dimensionality of the problem. Our approach uses the signal correlation in the $x$-$q$ space to help reduce the complexity of the problem by imposing that the reconstructed signal should be smooth in a non-local sense. That is, neighboring points in the product space of the signal space and the $x$-$q$ space should be reconstructed using similar values. Unlike the commonly used $x$-space regularization (Coupé et al., 2013), which takes into account spatial correlation, $x$-$q$ space regularization allows correlation across DW images collected using different gradient directions and strengths to be considered. This is fitting when considering the fact that WM structures might be curved and hence causing rapid changes within a DW image.

For simplicity, we represent the LAR data using vector $\mathbf{y}$ and the HAR data that we need to estimate using vector $\mathbf{x}$. The objective function of our problem is as follows:

$$\epsilon^2(\mathbf{x}) = \frac{\lambda}{2}\|\mathbf{A}\mathbf{x} - \mathbf{y}\|_2^2 + \frac{1}{4} \sum_{(i,k)\in\Omega} \sum_{(j,l)\in\mathcal{V}(i,k)} w[i,k;j,l]\|\mathbf{R}_{i,k}\mathbf{x} - \mathbf{R}_{j,l}\mathbf{x}\|_2^2, \qquad (1)$$

where $\mathbf{A}$ is a $q$-space downsampling operator, $w[i,k;j,l]$ is a weight determined by neighborhood matching, $\mathbf{R}_{i,k}$ is an operator that extracts the diffusion signal associated with an $x$-space index $i$ and a $q$-space index $k$. The penalty function consists of a data fidelity term and a regularization term based on $x$-$q$ space neighborhood matching. To minimize (1), we compute the derivative and equate it to zero:

$$0 = \frac{d\epsilon^2(\mathbf{x})}{d\mathbf{x}} = \lambda \mathbf{A}^\top(\mathbf{A}\mathbf{x} - \mathbf{y}) + \frac{1}{2} \sum_{(i,k)\in\Omega} \sum_{(j,l)\in\mathcal{V}(i,k)} w[i,k;j,l](\mathbf{R}_{i,k} - \mathbf{R}_{j,l})^\top(\mathbf{R}_{i,k} - \mathbf{R}_{j,l})\mathbf{x}$$

$$= \lambda \mathbf{A}^\top(\mathbf{A}\mathbf{x} - \mathbf{y}) + \frac{1}{2} \sum_{(i,k)\in\Omega} \sum_{(j,l)\in\mathcal{V}(i,k)} w[i,k;j,l]\mathbf{R}_{i,k}^\top \mathbf{R}_{i,k}\mathbf{x}$$

$$- \frac{1}{2} \sum_{(i,k)\in\Omega} \sum_{(j,l)\in\mathcal{V}(i,k)} w[i,k;j,l]\mathbf{R}_{i,k}^\top \mathbf{R}_{j,l}\mathbf{x} - \frac{1}{2} \sum_{(i,k)\in\Omega} \sum_{(j,l)\in\mathcal{V}(i,k)} w[i,k;j,l]\mathbf{R}_{j,l}^\top \mathbf{R}_{i,k}\mathbf{x}$$

$$+ \frac{1}{2} \sum_{(i,k)\in\Omega} \sum_{(j,l)\in\mathcal{V}(i,k)} w[i,k;j,l]\mathbf{R}_{j,l}^\top \mathbf{R}_{j,l}\mathbf{x}. \quad (2)$$

Based on the facts that the neighborhood is symmetric (i.e., if $(j,l) \in \mathcal{V}(i,k)$, then $(i,k) \in \mathcal{V}(j,l)$) and the weights are symmetric (i.e., $w[i,k;j,l] = w[j,l;i,k]$) (Protter et al., 2009), we have

$$\sum_{(i,k)\in\Omega} \sum_{(j,l)\in\mathcal{V}(i,k)} w[i,k;j,l]\mathbf{R}_{i,k}^\top \mathbf{R}_{i,k}\mathbf{x} = \sum_{(i,k)\in\Omega} \sum_{(j,l)\in\mathcal{V}(i,k)} w[i,k;j,l]\mathbf{R}_{j,l}^\top \mathbf{R}_{j,l}\mathbf{x}$$

$$\sum_{(i,k)\in\Omega} \sum_{(j,l)\in\mathcal{V}(i,k)} w[i,k;j,l]\mathbf{R}_{j,l}^\top \mathbf{R}_{i,k}\mathbf{x} = \sum_{(i,k)\in\Omega} \sum_{(j,l)\in\mathcal{V}(i,k)} w[i,k;j,l]\mathbf{R}_{i,k}^\top \mathbf{R}_{j,l}\mathbf{x}. \quad (3)$$

Equation (2) can be simplified using (3), giving

$$0 = \lambda \mathbf{A}^\top(\mathbf{A}\mathbf{x} - \mathbf{y}) + \sum_{(i,k)\in\Omega} \sum_{(j,l)\in\mathcal{V}(i,k)} w[i,k;j,l]\mathbf{R}_{i,k}^\top \mathbf{R}_{i,k}\mathbf{x} - \sum_{(i,k)\in\Omega} \sum_{(j,l)\in\mathcal{V}(i,k)} w[i,k;j,l]\mathbf{R}_{i,k}^\top \mathbf{R}_{j,l}\mathbf{x}. \quad (4)$$

Equation (4) can be solved directly but involves the inversion of a very large matrix, therefore we choose instead to use fixed-point iteration to solve the problem, as suggested in (Protter et al., 2009). If we let $\mathbf{x}^n$ be the solution at iteration $n$, the following can be proven to be convergent (Protter et al., 2009):

$$0 = \lambda \mathbf{A}^\top(\mathbf{A}\mathbf{x}^n - \mathbf{y}) + \sum_{(i,k)\in\Omega} \sum_{(j,l)\in\mathcal{V}(i,k)} w[i,k;j,l]\mathbf{R}_{i,k}^\top \mathbf{R}_{i,k}\mathbf{x}^n - \sum_{(i,k)\in\Omega} \sum_{(j,l)\in\mathcal{V}(i,k)} w[i,k;j,l]\mathbf{R}_{i,k}^\top \mathbf{R}_{j,l}\mathbf{x}^{n-1}. \quad (5)$$

The solution $\mathbf{x}$ can be obtained iteratively using

$$\mathbf{x}^n = \left( \lambda \mathbf{A}^\top \mathbf{A} + \sum_{(i,k)\in\Omega} \sum_{(j,l)\in\mathcal{V}(i,k)} w[i,k;j,l]\mathbf{R}_{i,k}^\top \mathbf{R}_{i,k} \right)^{-1}$$

$$\times \left( \lambda \mathbf{A}^\top \mathbf{y} + \sum_{(i,k)\in\Omega} \sum_{(j,l)\in\mathcal{V}(i,k)} w[i,k;j,l]\mathbf{R}_{i,k}^\top \mathbf{R}_{j,l}\mathbf{x}^{n-1} \right). \quad (6)$$

The matrix inversion in (6) involves a block diagonal matrix and can therefore be done effectively.

## 2.3 Implementation Issues

### 2.3.1 Initialization

The data are transformed so that the noise is Gaussian distributed as described in (Koay et al., 2009). The algorithm is then initialized using an upsampled version of $\mathbf{y}$, which is obtained via interpolation using SHs.

### 2.3.2 Neighborhood matching

Neighborhood matching is performed based on the upsampled version of $\mathbf{y}$. The resulting weights remain unchanged until a solution $\mathbf{x}$ is obtained. In principle, we can use $\mathbf{x}$ to re-estimate the weights and rerun the algorithm to obtain a refined solution. However, our experimental results indicate that the benefit of doing so is minimal. Therefore, we will only show results without weight re-estimation.

### 2.3.3 Stopping criterion

We stop the algorithm when the mean absolute difference (MAD) between the outcomes of two iterations, i.e., $\mathbf{x}^{n-1}$ and $\mathbf{x}^n$, is less than a constant `tol`. We define `tol` $= \beta \sigma_G$, where $\sigma_G$ is the standard derivation of the Gaussian noise and $\beta$ is a constant.

## 3 EXPERIMENTS

We evaluated the proposed $q$-space upsampling algorithm using both synthetic and real data. Through grid search, we found that $\lambda = 100$ and $\beta = 10^{-3}$ give the best results. We have two comparison baselines, SRBF interpolation and SH interpolation.

### 3.1 Datasets

#### 3.1.1 Synthetic Data

A set of synthetic data was generated using phantom$\alpha$s (Caruyer et al., 2014). Two sets of diffusion gradient directions were utilized to simulate the HAR and LAR DMRI data. The LAR gradients were generated by dividing the faces of an icosahedron three times and discarding antipodal symmetric directions, giving us a total of 81 directions uniformly distributed on a hemisphere. Based on the same strategy, we generated the HAR gradient directions by dividing the faces of the icosahedron four times to obtain 321 directions. Data for 3 shells, $b = 1000, 2000, 3000 \, \text{s/mm}^2$, were generated. Finally, we added four levels (SNR = 15, 20, 25, 30) of 32-channel noncentral chi (nc-$\chi$) noise to the LAR data.

#### 3.1.2 Real Data

DMRI data were acquired for three infants at three different time points: 0 month, 6 months, and 12 months. All enrolled subjects had informed consent provided by parent/guardian. The experimental protocols were approved by the Institutional Review Board of the University of North Carolina (UNC) School of Medicine. The study was carried out in accordance with the approved guidelines. All the data are acquired using a Siemens 3T Magnetom Prisma MR scanner and a standard imaging protocol: $140 \times 140$ imaging matrix, $1.5 \times 1.5 \times 1.5 \, \text{mm}^3$ resolution, TE=$88 \, \text{ms}$, TR=$2{,}365 \, \text{ms}$, 32-channel receiver coil, $b = 700, 1500, 3000 \, \text{s/mm}^2$, and 144 non-collinear gradient directions. We uniformly selected 72 gradient directions to generate the LAR data for evaluation.

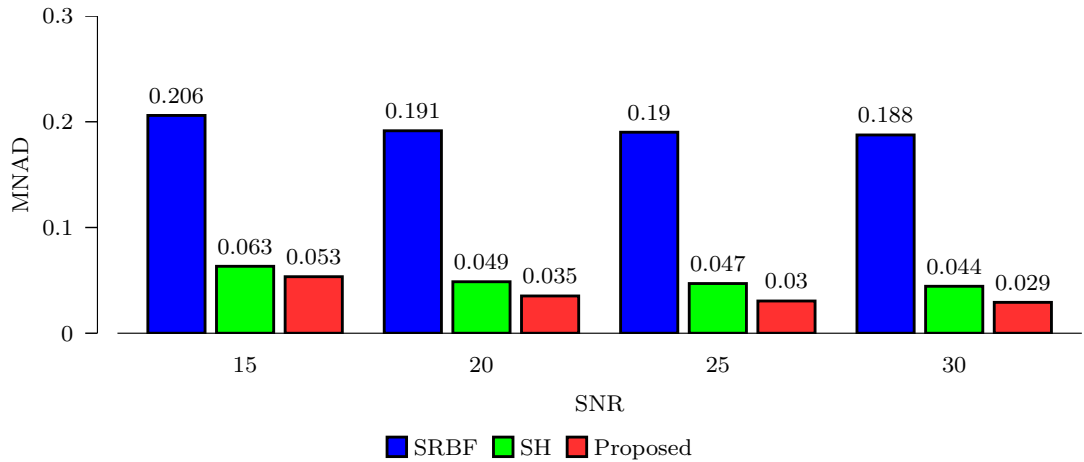

**Figure 2. MNAD Comparison – Synthetic Data.** Quantitative evaluation using synthetic data via MNAD of FA images.

## 3.2 Evaluation Methods

Quantitative and qualitative evaluations were performed as described in the following:

1. **RMSE maps:** We computed voxel-wise RMSE value between two sets of DMRI datasets to measure their similarity locally.
2. **FA images:** We computed the FA images using the iterative weighted tensor fitting method presented in (Salvador et al., 2005).
3. **Absolute Difference (AD) Maps:** We compute the AD map between one FA image and the ground truth FA image to evaluate the performance of the algorithm locally.
4. **Mean normalized absolute difference (MNAD):** We computed MNAD value by 1) computing AD map, 2) normalizing the AD map using the ground truth FA image voxel-wisely, and 3) computing the mean value of the normalized map within the brain region.
5. **Fiber ODFs:** We compute fiber ODFs using the method presented in (Yap et al., 2016b), which caters to multiple tissue types using multi-shell data.

## 3.3 Results

### 3.3.1 MNAD Comparison – Synthetic Data

Using the FA image of noise-free HAR data as ground truth, we evaluated the quality of the upsampled data using MNAD. The results, shown in Fig. 2, indicate that the proposed method outperforms SRBF interpolation and SH interpolation for all noise levels. The largest improvement over the second best method, SH interpolation, is 0.017 when SNR = 25.

### 3.3.2 DW Images – Synthetic Data

The full views and close-up views of DW images, shown in the top two rows of Fig. 3, indicate that the proposed method results in better structural contrast. The close-up views of RMSE maps, shown in the bottom row of Fig. 3, indicate that our method gives lower RMSE than SRBF interpolation and SH interpolation, which demonstrates that the upsampled data given by our method is closer to ground truth.

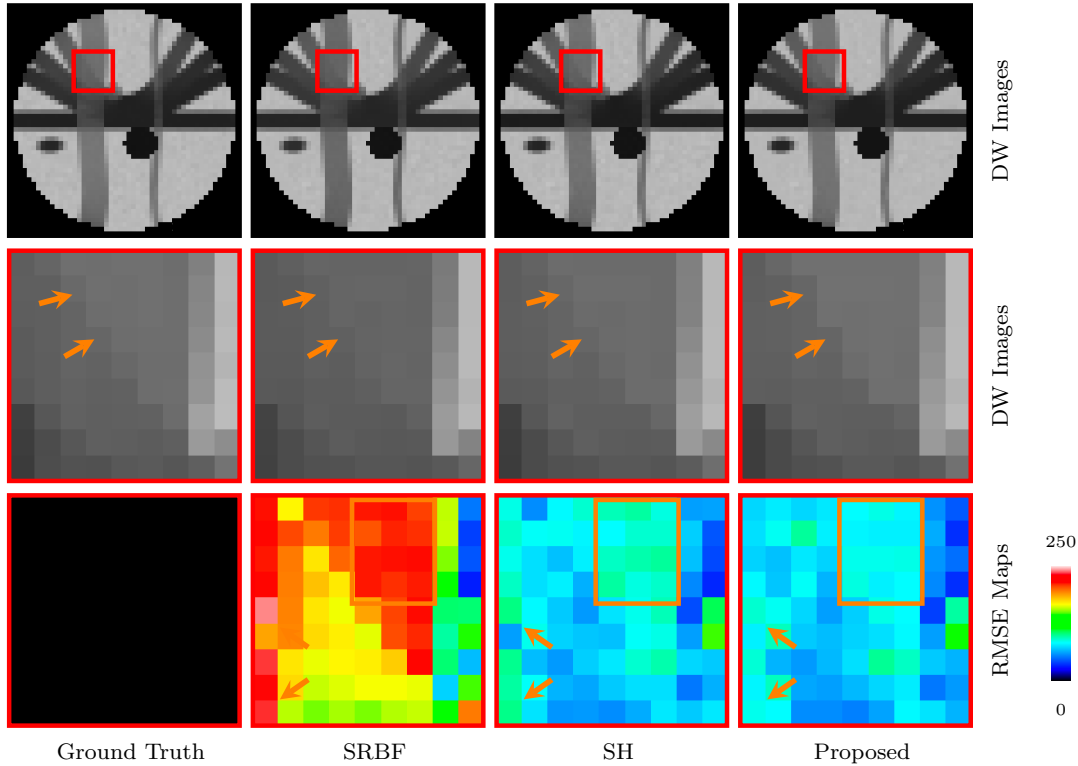

**Figure 3. DW Images and RMSE Maps – Synthetic Data.** Comparison of upsampling results for $b = 1000\,\text{s/mm}^2$ and nc-$\chi$ noise (SNR = 30).

### 3.3.3 FA Images – Synthetic Data

The top row of Fig. 4 shows the FA images given by ground truth data and upsampled data. We use warm colors to represent large FA values for a better visualization. Compared with SRBF interpolation and SH interpolation, our method produces an FA image closer to the ground truth. This observation is further confirmed by the AD maps of the FA images, shown in the bottom row of Fig. 4. Our method yields lower AD values, indicating better performance.

### 3.3.4 Fiber ODFs – Synthetic Data

Accurate ODF estimation relies on sufficient angular samples. The ODFs, shown in Fig. 5, indicate that our method gives clean and coherent ODFs that are close to the ground truth. In contrast, spurious peaks are introduced by SRBF interpolation and SH interpolation.

### 3.3.5 Diffusion Signal Profiles – Synthetic Data

For a more direct visualization of the upsampled data, we rendered the signal values on a sphere. The results, shown in Fig. 6, indicate that our method gives values that are close to the ground truth.

### 3.3.6 MNAD Comparison – Real Data

We also computed MNAD values for the quantitative evaluation of real data experimental results. Fig. 7 shows the MNAD between the FA images given by the upsampled data and the original HAR data. For all time points, our method outperforms SRBF interpolation and SH interpolation, with a largest MNAD reduction of 0.026 over the second best method at 12-months time point.

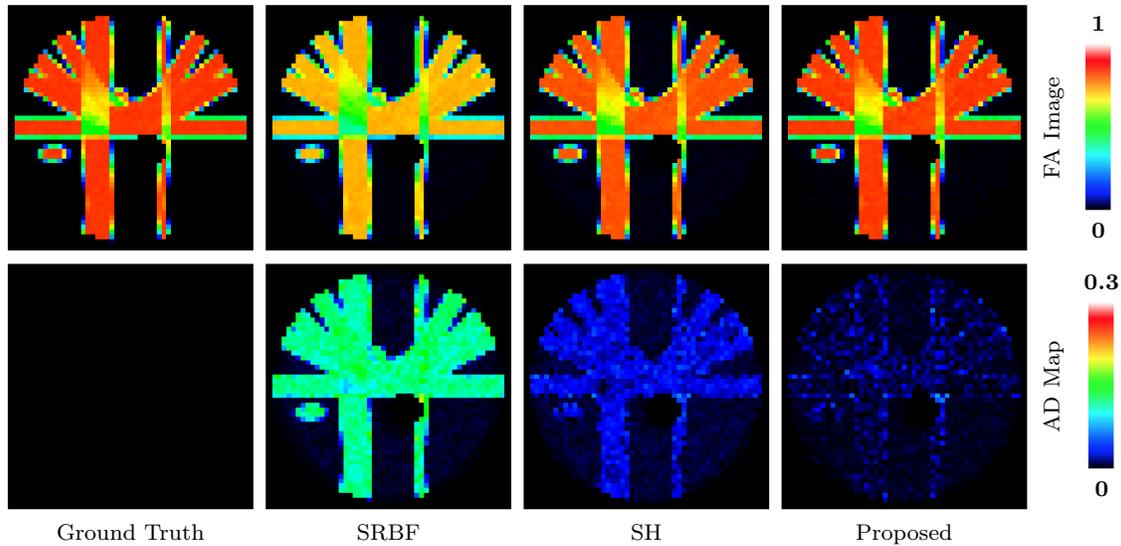

**Figure 4. FA Images and AD Maps – Synthetic Data.** Evaluation of accuracy in terms of FA using synthetic dataset with nc-$\chi$ noise (SNR = 30). The color of FA images represents the FA value, e.g., warmer color means a larger FA value.

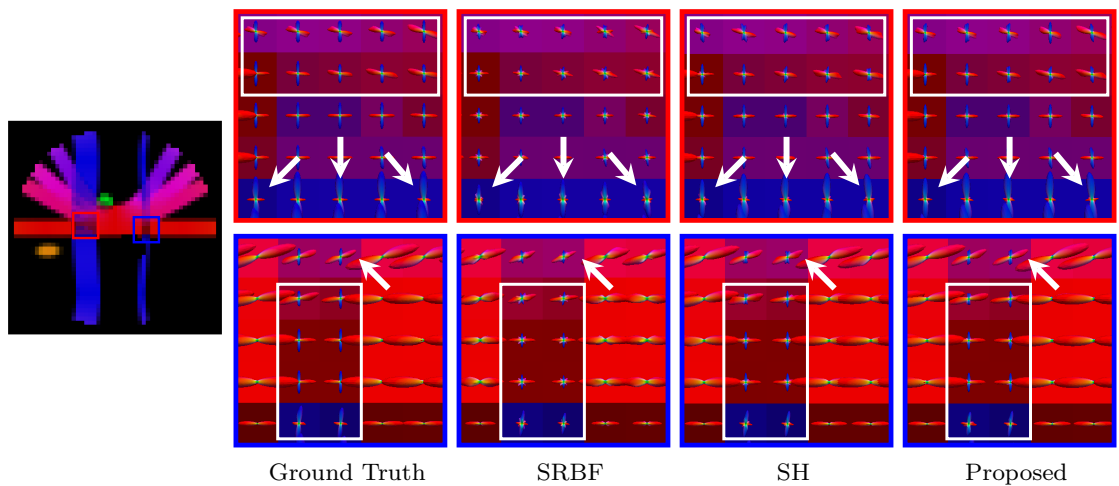

**Figure 5. Fiber ODFs – Synthetic Data.** Fiber ODF comparison using synthetic data with nc-$\chi$ noise (SNR = 30).

### 3.3.7 DW Images – Real Data

The observations from Fig. 8 and 9 for real data are consistent with that in Fig. 3. The DW image given by our method shows more subtle structural details and is closer to the one in the original HAR data.

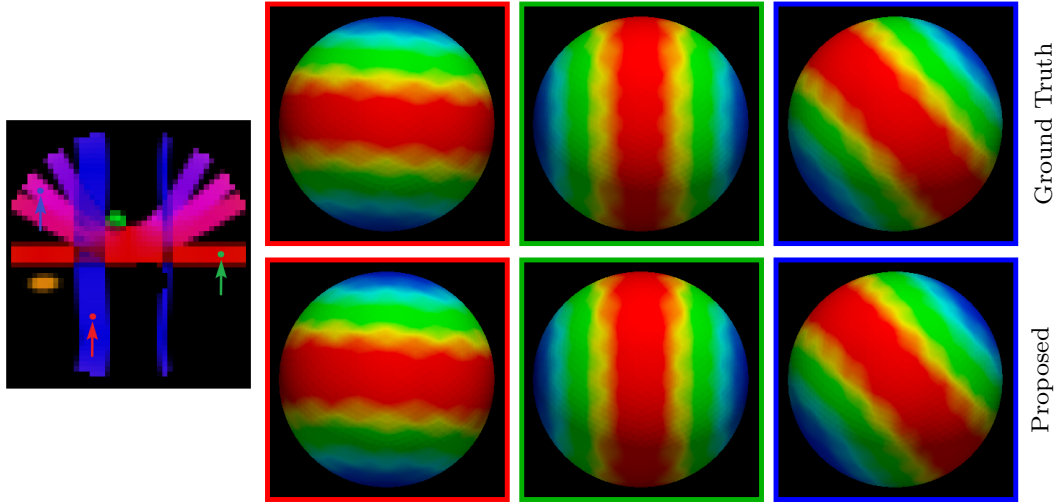

**Figure 6. Diffusion Signal Profiles – Synthetic Data.** The diffusion signals are rendered on a sphere for visual comparison. The colored FA image is shown on the far left for reference.

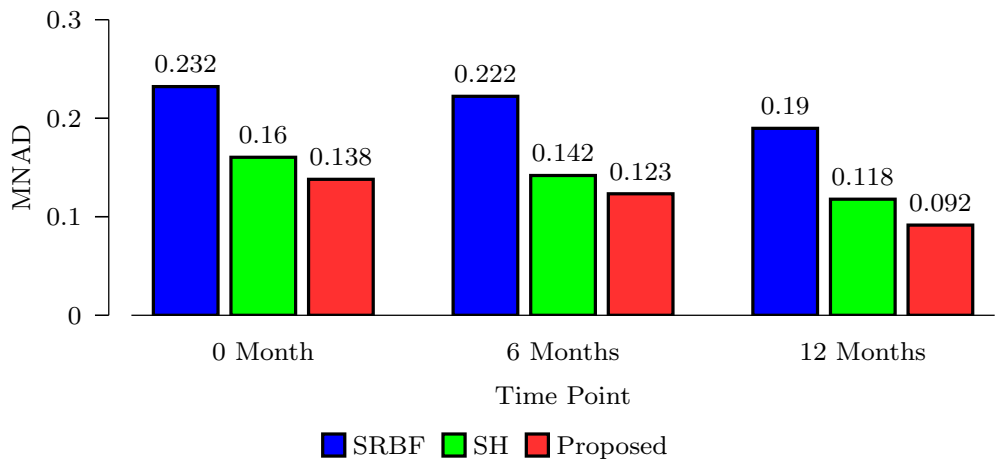

**Figure 7. MNAD Comparison – Real Data.** Quantitative evaluation using infant data via MNAD of FA images.

### 3.3.8 FA Images – Real Data

Fig. 10 and 11 further confirm our observation in Fig. 4. Our method produces low AD values and gives an FA image that is close to that given by the original HAR data.

### 3.3.9 Fiber ODFs – Real Data

Fig. 12 indicates that our method gives clean and coherent ODFs that are very similar to those given by the original HAR data. In contrast, SRBF interpolation and SH interpolation produce ODFs with a large number of spurious peaks.

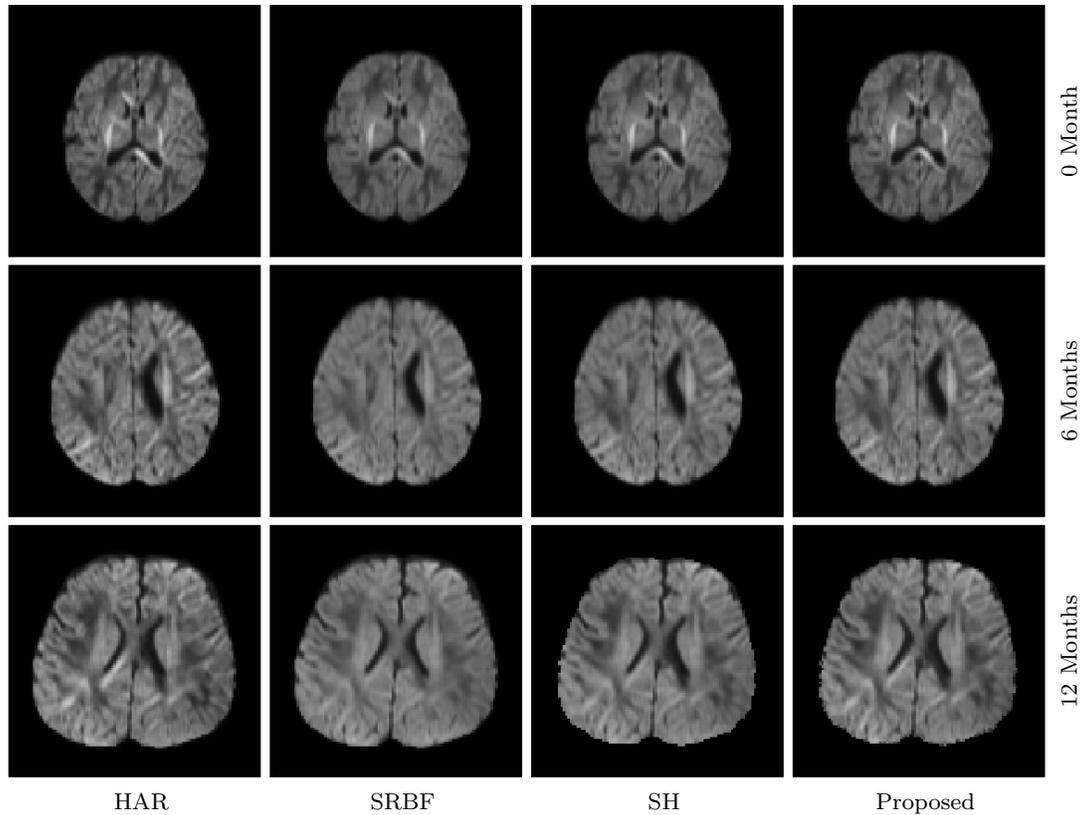

**Figure 8. DW Images – Real Data.** Comparison of DW images with $b = 1,500\,\text{s/mm}^2$.

## 4 DISCUSSIONS

Our method is effective because it preserves sharpness of signal profiles in $q$-space during upsampling. Utilizing non-local smoothness as prior, it takes into account signal similarity in $x$-$q$ space and avoids the pitfall of averaging over disparate signals. For the curved white matter structures, considering only signal correlation in $x$-space (i.e., a fixed point in $q$-space) is problematic because the signal changes rapidly across space. On the other hand, considering only signal correlation in $q$-space (i.e., a fixed point in $x$-space) causes smoothing of anisotropic signal profiles due to sharp changes across $q$-space measurements. Our method harnesses the fact that the signal is smooth in the joint $x$-$q$ space, even for highly curved structures.

## 5 CONCLUSION

We have presented a regularization framework for $q$-space upsampling. The relationships of signals in $x$-$q$ space are used to regularize the inverse problem associated with recovering the HAR DMRI data. Extensive experiments on synthetic and infant DMRI data indicate that our method is able to produce HAR DMRI data with significantly improved quality. Future research effort will be directed to extending the current framework for resolution enhancement in both $x$-space and $q$-space.

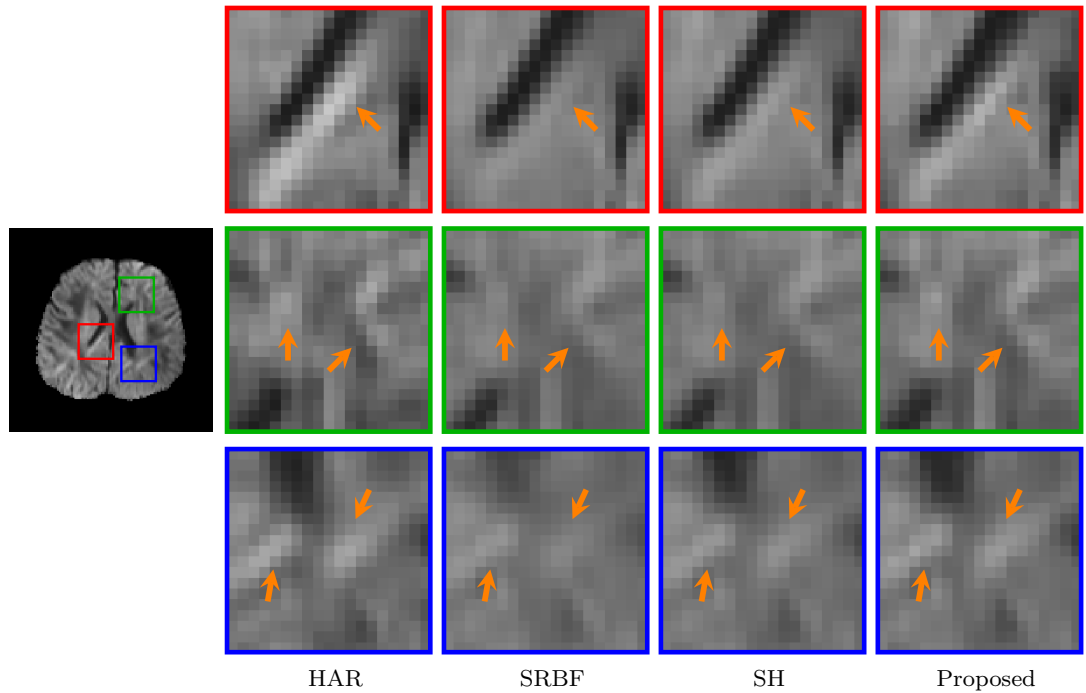

**Figure 9. Close-Up Views of DW Images – Real Data.** Regional close-up views of the DW images of a 12-month infant subject.

## CONFLICT OF INTEREST STATEMENT

The authors declare that the research was conducted in the absence of any commercial or financial relationships that could be construed as a potential conflict of interest.

## AUTHOR CONTRIBUTIONS

GC and P-TY implemented the code and designed the experiments. GC drafted the manuscript. P-TY revised the manuscript. WL provided the clinical data. BD, YZ and DS participated in idea discussion and reviewed the manuscript.

## FUNDING

This work was supported in part by NIH grants (NS093842, EB022880, EB006733, EB009634, AG041721, MH100217, and AA012388) and a NSFC grant (11671022).

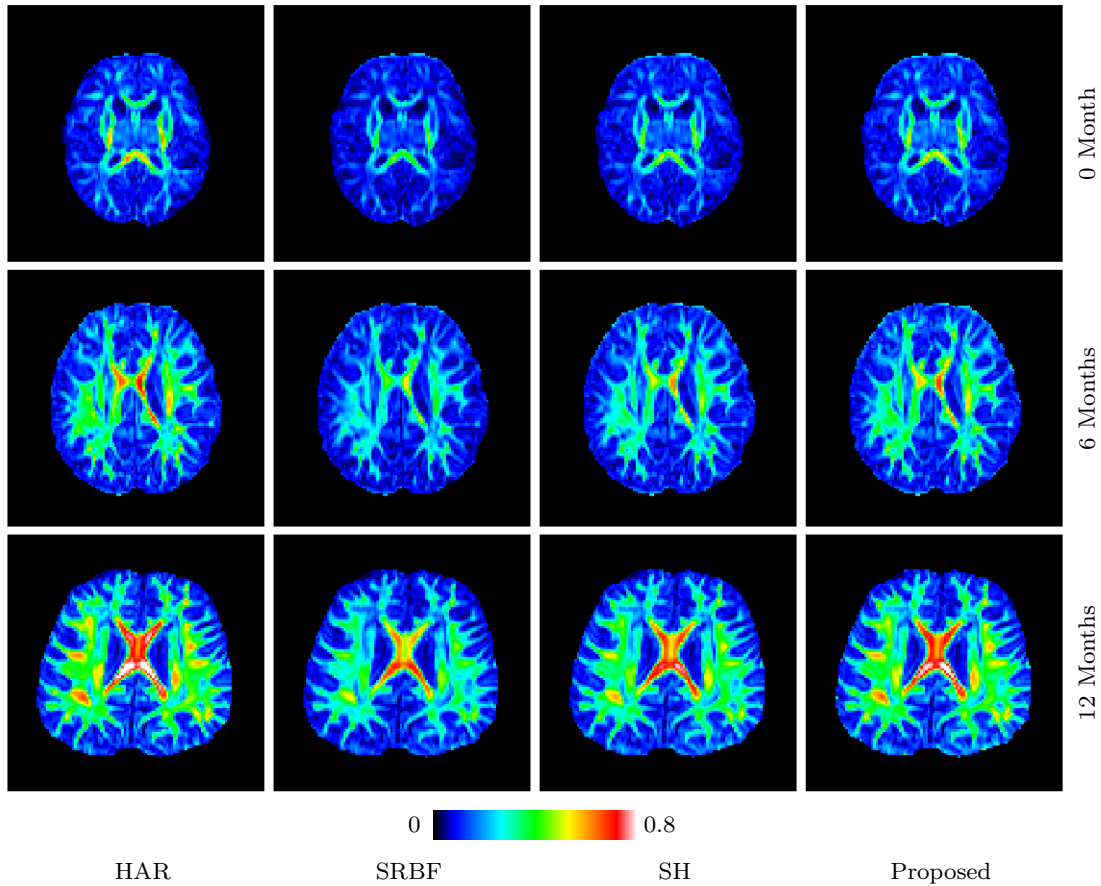

**Figure 10. FA Images – Real Data.** Comparison of FA images using infant data. The color of FA images represents the FA value, e.g., warmer color means a larger FA value.

Chen, G., Dong, B., Zhang, Y., Shen, D., and Yap, P.-T. (2017a). Denoising of diffusion MRI data using manifold neighborhood matching. In *The Organization for Human Brain Mapping (OHBM) Annual Meeting*

Chen, G., Dong, B., Zhang, Y., Shen, D., and Yap, P.-T. (2017b). $q$-space upsampling using $x$-$q$ space regularization. In *International Conference on Medical Image Computing and Computer-Assisted Intervention* (Springer), 620–628

Chen, G., Wu, Y., Shen, D., and Yap, P.-T. (2016). XQ-NLM: Denoising diffusion MRI data via $x$-$q$ space non-local patch matching. In *Medical Image Computing and Computer-Assisted Intervention*. vol. 9902, 587–595

Coupé, P., Manjón, J. V., Chamberland, M., Descoteaux, M., and Hiba, B. (2013). Collaborative patch-based super-resolution for diffusion-weighted images. *NeuroImage* 83, 245–261

Descoteaux, M., Angelino, E., Fitzgibbons, S., and Deriche, R. (2007). Regularized, fast, and robust analytical Q-ball imaging. *Magnetic resonance in medicine* 58, 497–510

Dong, B. (2017). Sparse representation on graphs by tight wavelet frames and applications. *Applied and Computational Harmonic Analysis* 42, 452–479

Dubois, J., Dehaene-Lambertz, G., Kulikova, S., Poupon, C., Hüppi, P. S., and Hertz-Pannier, L. (2014). The early development of brain white matter: a review of imaging studies in fetuses, newborns and infants. *Neuroscience* 276, 48–71

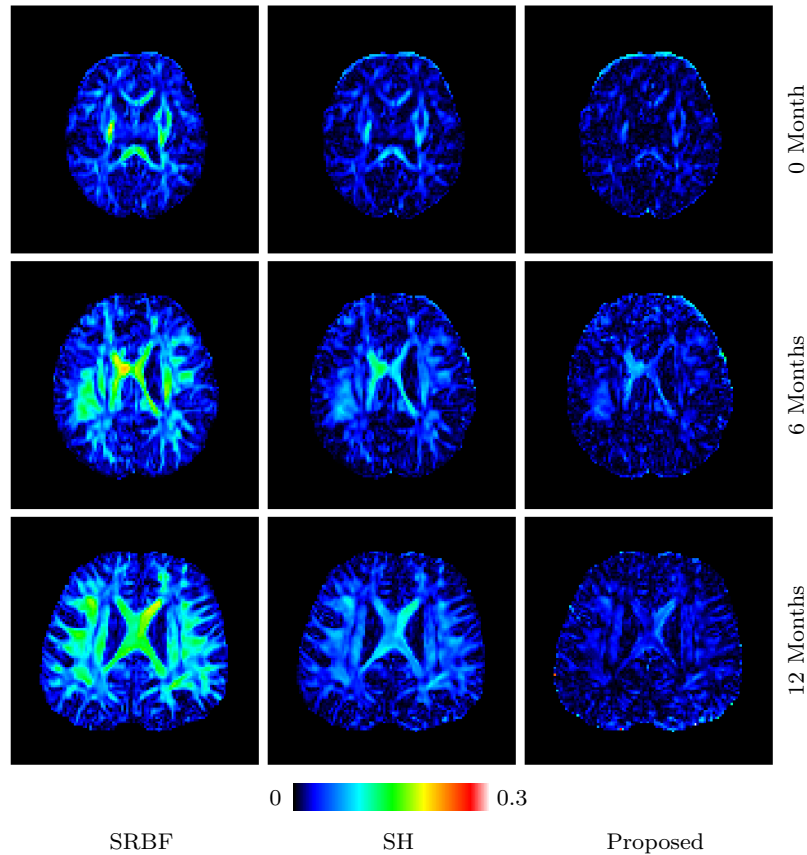

**Figure 11. Absolute Difference Maps – Real Data.** Comparison of absolute difference maps of FA images using infant data.

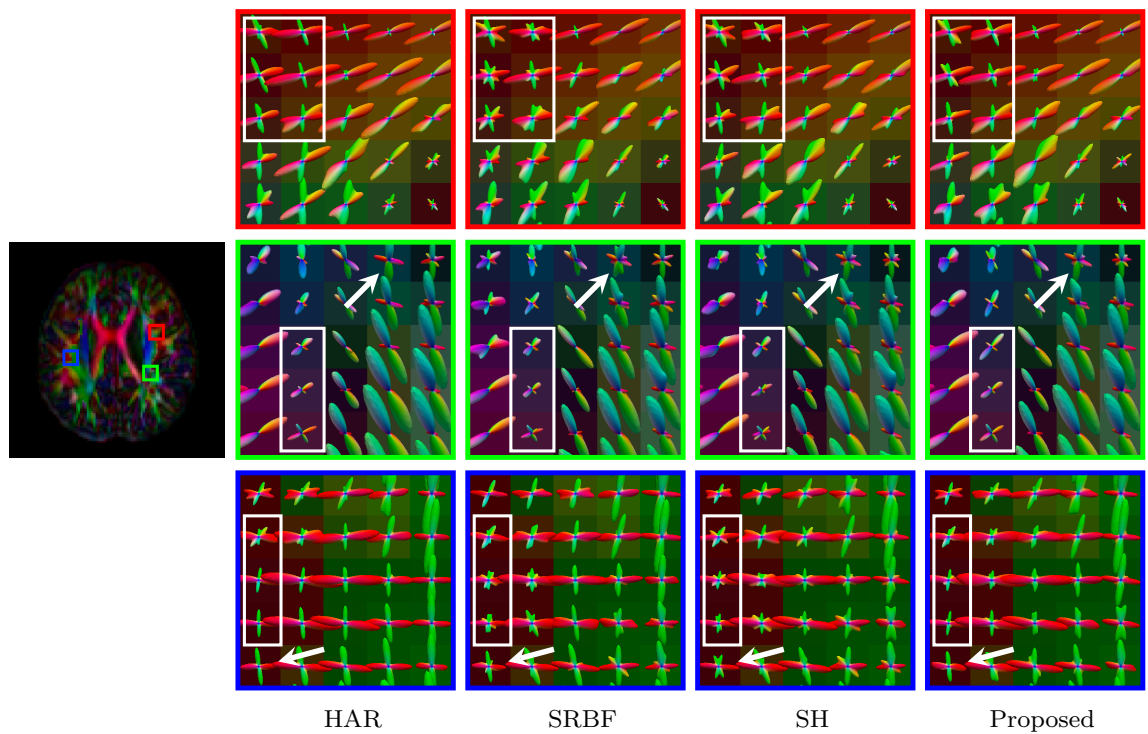

HAR     SRBF     SH     Proposed

**Figure 12. Fiber ODFs – Real Data.** Comparison of fiber ODFs using the DMRI data of a 6-Month infant subject.